# Hermite-Gaussian-mode coherently composed states and deep learning based free-space optical communication link

ZILONG ZHANG,[1,2,3,*] SUYI ZHAO,[1,2,3] WEI HE,[1,2,3] YUAN GAO,[1,2,3] XIN WANG,[1,2,3] YUCHEN JIE,[1,2,3] XIAOTIAN LI,[1,2,3] YUQI WANG,[1,2,3] AND CHANGMING ZHAO[1,2,3]

[1] School of Optics and Photonics, Beijing Institute of Technology, 5 South Zhongguancun Street, Beijing 100081, China
[2] Key Laboratory of Photoelectronic Imaging Technology and System (Beijing Institute of Technology), Ministry of Education, Beijing 100081, China
[3] Key Laboratory of Photonics Information Technology (Beijing Institute of Technology), Ministry of Industry and Information Technology, Beijing 100081, China
*zlzhang@bit.edu.cn

**Abstract:** In laser-based free-space optical communication, besides OAM beams, Hermite-Gaussian (HG) modes or HG-mode coherently composed states (HG-MCCS) can also be adopted as the information carrier to extend the channel capacity with the spatial pattern based encoding and decoding link. The light field of HG-MCCS is mainly determined by three independent parameters, including indexes of HG modes, relative initial phases between two eigenmodes, and scale coefficients of the eigenmodes, which can obtain a large number of effective coding modes at a low mode order. The beam intensity distributions of the HG-MCCSs have obvious distinguishable spatial characteristics and can keep propagation invariance, which are convenient to be decoded by the convolutional neural network (CNN) based image recognition method. We experimentally utilize HG-MCCS to realize a communication link including encoding, transmission under atmospheric turbulence (AT), and decoding based on CNN. With the index order of eigenmodes within six, 125 HG-MCCS are generated and used for information encoding, and the average recognition accuracy reached 99.5% for non-AT conditions. For the 125-level color images transmission, the error rate of the system is less than 1.8% even under the weak AT condition. Our work provides a useful basis for the future combination of dense data communication and artificial intelligence technology.

## 1. Introduction

Structured laser beams-based free-space optics (FSO) provides a new strategy for improving channel capacity in laser communicaitons. The information carried by spatial amplitude, phase, and polarization of the structured laser beams can all be used in FSO. As the most studied structured laser beam, orbital angular momentum (OAM) beam was widely applied. There are generally two ways of utilizing OAM in FSO [1]. One is mode division multiplexing (MDM), belonging to the category of space division multiplexing [2-4], in which different OAM beams as independent data channels propagating over the same spatial medium simultaneously. The other is the mode encoding system [5-6], which maps the quasi-modulated information to the OAM mode sequence in one-to-one correspondence, and has important value in the field of quantum communication. The number of data bits that can be carried by a single OAM state is $\log_2 N$, where N represents the number of orthogonal coding states [7-8]. However, laser beam propagation will face a series of challenges in FSO links, including atmospheric turbulence, pointing misalignment, and beam divergence, and the OAM beam as the information carrier will aggravate these problems [9].

At the same time, in order to increase the number of channels or coding modes, higher-order OAM modes are required. However, the divergence angle increases rapidly with the increase of mode order, which not only puts forward higher requirements for the optical collimation system but also has a larger beam radius after collimation [10]. The beam with a larger radius requires a larger receiver aperture and is easier to be disturbed by atmospheric turbulence [11,12]. The aperture limitation of the receiving system and the misalignment caused by the beam drift are the two most prominent problems in FSO, which will cause power loss and mode crosstalk [13-14]. In fact, in addition to OAM beams, other beams with orthogonality can also be used as carriers in FSO, such as the HG beam [15] and multi-vortex geometric beams (MVGBs) [16]. HG beams have stronger anti-crosstalk ability and lower power loss than LG beams under certain conditions [17]. The MVGB has higher divergence degeneracy and lower error rates caused by center offset and coherent background noise than OAM beams [16].

If only lower mode orders are used and more data capacity can be realized under a smaller transmission aperture, which will effectively alleviate the above problems. Here the coherent superposition states between basic eigenmodes provide a possible solution. Although the mode orthogonality of each channel is required to avoid crosstalk in MDM, it is only necessary to distinguish the received mode in the mode encoding system. Because different coding modes transmit in the form of pulses and the information reaching the receiver at different times is unique. Therefore, eigenmodes coherently composed beams can be used as information carriers [18,19].

Moreover, if taking HG modes as basic modes to generate HG-MCCS, it will produce more abundant coherently composed states than that with LG modes. Firstly, the $HG_{mn}$ beam has more distinguishable degenerate patterns than the $LG_{pl}$ beam with the same mode order $N=m+n=2p+|l|$, due to LG modes with a pair of opposite topological charge $l$ having the same light intensity distribution. Secondly, since the HG mode is under Cartesian symmetry, the differences in spatial amplitude of the HG-MCCS caused by the phase difference between basic component modes are very obvious. While, for the LG mode with rotational symmetry, the spatial pattern of the LG-MCCS will only rotate by a certain angle depending on the phase difference between the component LG modes, which is difficult to be distinguished. Considering the number of efficient degenerate eigenmodes and their combinations with different phases, HG-MCCS can obtain a higher coding capacity wihin a lower mode order. The lower mode order means a smaller beam radius under the same beam divergence, which can weaken the influence of AT and reduce the requirements on the optical system, thus reducing mode crosstalk and power loss.

Another convenience with HG-MCCS is the unique and abundant intensity distributions can be easily decoded by image-based recognition. Based on the orthogonality between modes, the main traditional decoding method is to decompose LG [20-26] and HG [27] modes by conjugate mode sorting. However, with the promotion of deep learning, more and more researches on laser transverse mode recognition based on image classifications were presented recently, including both OAM beams [28-33] and HG beams [34,35]. In 2019, an 18-layer RESNET variant CNN was designed to detect 21 HG modes with an accuracy of more than 99% [34]. In 2020, CNN was used as a regression tool to combine classification training with the modal decomposition of HG beams for the first time [35]. The above works provide a good exploration experience for the deep learning recognition of HG-MCCS.

In this paper, the FSO link research of encoding, transmission, and deep learning decoding of HG-MCCS is carried out. A variety of distinguishable HG-MCCS are generated by modulation of DMD loaded with holograms, which constitute the coding modes set. With the basic transverse mode order $N \leq 6$, 125 effective HG-MCCS are selected for encoding. The disturbance of AT through long-distance beam propagation is also simulated by the same DMD that generates the HG-MCCS beams above. The intensity distribution of the transmitted beam is collected with a CCD camera and recognized by deep learning method. Through experimental verification, high accuracy transmission of 50×50 pixels color images is achieved based on the above link. The mode recognition bit error rates of full 125-level color images under the condition of no turbulence, 1 km, and 3 km weak turbulence are lower than 0.5%, 2%, and 6% respectively.

## 2. Theory model

*2.1 HG-MCCS with three independent parameters*

The mathematical expression of a single HG mode is well known as below,

$$HG_{mn}(x,y,z) = \frac{C_{m,n}^{HG}}{\omega^2}\exp(-\frac{x^2+y^2}{\omega^2})H_m(\frac{\sqrt{2}x}{\omega})H_n(\frac{\sqrt{2}y}{\omega})\exp\left[ikz + ik\frac{x^2+y^2}{2R(z)} - i(m+n+1)\Psi(z)\right] \quad (1)$$

Where $C_{m,n}^{HG}$ is the normalized constant of the HG mode, $H_m$ and $H_n$ are the $m$-th and $n$-th order Hermite polynomial. $\omega$ is the half-width of the beam at the position z, $\omega^2 = \omega_0^2(z_R^2+z^2)/z_R^2$, $\omega_0$ is the waist radius of the fundamental mode beam, $z_R$ and is the Rayleigh length $R(z)$ is the radius of curvature associated with $z$ and $\Psi(z) = \arctan(z/z_R)$ is the Gouy phase.

The electrical field of the propagation-invariant HG-MCCS is the coherent superposition of HG modes with the same mode order, and the mathematical expression can be derived as,

$$E(x,y,z) = \sum_i \alpha_i HG_{m_i,n_i}(x,y,z)\exp(i\Delta\varphi_i) \quad (2)$$

The item $\exp(i\Delta\varphi_i)$ is the phase difference between the $i^{th}$ and $1^{st}$ HG mode in the HG-MCCS. The normalization coefficient $\alpha_i$ is the relative ratio of each HG mode. $m_i+n_i$ are equal positive integers. It can be seen from Eq. (2) that the final electrical field is mainly determined by three parameters, namely the indexes of HG modes, the relative initial phases $\exp(i\Delta\varphi_i)$, and scale coefficients between modes $\alpha_i$. By changing the three parameters, various of different electrical fields, as well as intensity distributions, can be obtained. Fig. 1 shows some examples of intensity distributions of HG modes and LG modes coherently composed beams by changing the three parameters respectively.

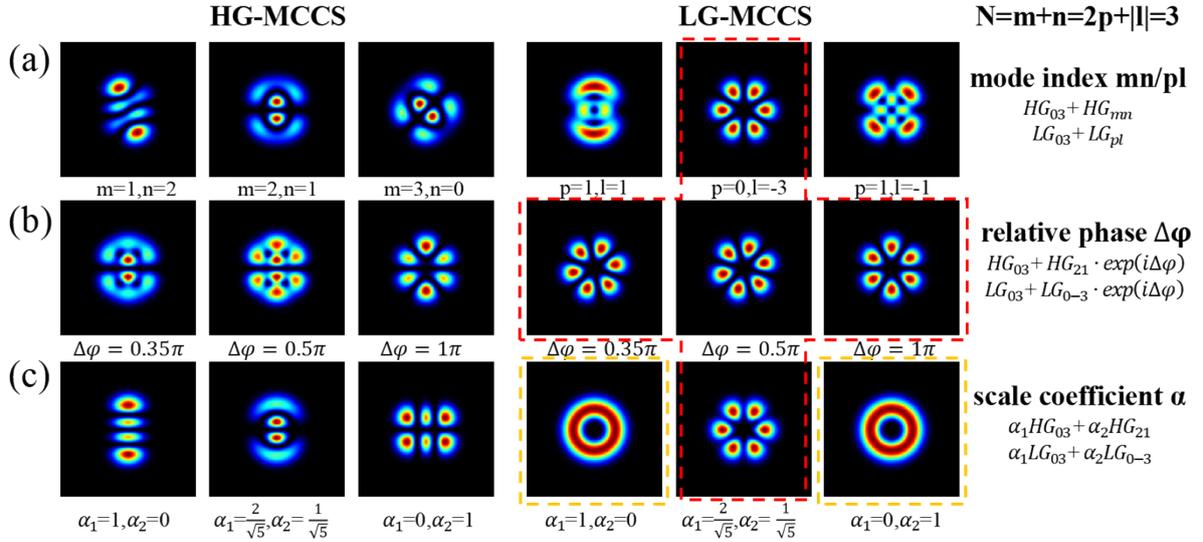

Fig. 1. The beam intensity distributions of HG-MCCS (left) and LG-MCCS (right) when mode order N=3. (a) Variations of beam intensity distributions by individually changing indexes of HG eigenmodes. (b) Variations of beam intensity distributions by individually changing relative initial phases between two eigenmodes. (c) Variations of beam intensity distributions by individually changing scale coefficients of two eigenmodes.

A variety of beam patterns can be obtained by coherently composing only a few low-order HG modes with different initial phases as shown in Fig. 1. This makes an obvious difference from the coherent superposition of LG modes, beam patterns of which rotating periodically with the change of phase difference [36,37] (marked by the red dotted box). What's more, even if the component proportions of the LG modes changed, the beam patterns will not show too obvious differences (marked by the orange dotted box). This prevents the LG-MCCS from adding much more effective encoding states than a single LG beam and therefore requires higher-order modes to increase the information capacity. However, too high order modes are not easy to generate efficiently and will meet a series of practical problems, such as diverging severely during transmission. Here, HG-MCCS offers us an opportunity to largely extend the encoding capacity with the several lowest orders of transverse modes.

The beam quality factor M² is further used to evaluate the propagation dynamics of eigenmode coherently composed beams in free space, the M² of HG-MCCS and LG-MCCS are expressed by Eq. (3) and Eq. (4) respectively. And the divergence of beams can be characterized by the half divergence angle θ.

$$M^2_{HGMCCB} = \sum_i (m_i + n_i + 1)|\alpha_i|^2 \tag{3}$$

$$M^2_{LGMCCB} = \sum_i (2p_i + |l_i| + 1)|\alpha_i|^2 \tag{4}$$

$$\theta = M^2 \lambda / \pi \omega_0 \tag{5}$$

A lower mode order means a lower M² value, which in turn leads to a smaller divergence angle and a smaller beam radius transmitted to the receiving end under the same conditions. At the same time, a larger number of coherently composed beams with distinguishable light intensity distribution with the same M² value means a higher divergence degeneracy of encoding modes. For example, more than thirty effective coding modes can be achieved by HG-MCCS when M²=4, compared with fewer than ten for LG-MCCS.

*2.2 HG-MCCS generation and turbulence simulation*

DMD is one of the space modulators that can generate structured beams by the computer holography method. Though its efficiency is not as high as that of the SLM, it can generate perfect beams and realize high-speed modulation due to its high spatial resolution and fast refresh speed. There are two basic hologram algorithms for generating structured beams with DMD, one is Binary Lee [38,39] and the other is Superpixel [40,41]. The Lee method modulates the single micromirror pixel independently and encodes the phase by the diffraction effect of irregular gratings. While the superpixel method combines N × N micromirrors into a single superpixel and modulates the phase by adding an extra off-axis to the imaging system. The basic principles of these two algorithms are similar, which can be concluded as a reverse calculation method. The mathematical formula for the transmittance of holograms can be expressed as below [42],

$$T(x,y) = \frac{1}{2} + \frac{1}{2}\text{sgn}\left\{\cos\left[\frac{2\pi x}{x_0} + \pi p(x,y)\right] - \cos[\pi \omega(x,y)]\right\} \tag{6}$$

Where $\text{sgn}(x)$ is the sign function, $x_0$ is the grating period, the slowly varying terms $\omega(x,y)$ and $p(x,y)$ are calculated by the target light field $S(x,y)$:

$$\omega(x,y) = \frac{1}{\pi}\arcsin\left[|S(x,y)|\right] \tag{7}$$

$$p(x,y) = \frac{1}{\pi}\text{ang}[S(x,y)] \tag{8}$$

The targeted beam can be obtained by illuminating the collimated Gaussian beam onto the DMD array and spatially filtering the +1st order diffraction beam out. The generated HG-MCCS is collimated and transmitted to a far distance to be received. Since the HG-MCCS adopted is propagation-invariant, only the influence of turbulence on the beam intensity distribution should be considered during the propagation.

In laboratory experiments, phase screens are usually inserted into the optical path to simulate the influence of AT. The phase disturbance is achieved by depicting the random phase screen according to the appropriate turbulent power spectrum and using devices such as SLM or turbulent plate [43]. In this work, the DMD is used to generate the HG-MCCS as well as simulate distortions induced by the turbulence simultaneously, which makes the experimental system more compact and chipper. The beam transmission model is established by equispaced random phase screens to simulate turbulence equivalently. The atmospheric turbulence phase screen is generated by the power spectrum inversion method as shown in Fig. 2(a), and the Von Karman model is selected to characterize the refractive index power spectral density as expressed by Eq. (9) [44].

$$\varphi(x,y) = \int_{-\infty}^{+\infty}\int_{-\infty}^{+\infty} R(\kappa_x,\kappa_y)\sqrt{2\pi^2 k^2 0.033 C_n^2 (\kappa^2+\kappa_0^2)^{\frac{-11}{6}} \exp\left[-\left(\frac{\kappa}{\kappa_m}\right)^2\right]\Delta z} \exp\left[i2\pi(\kappa_x x+\kappa_y y)\right] d\kappa_x d\kappa_y \quad (9)$$

Where $k$ is the wave number, $\Delta z$ is the transmission distance of each segment of the beam, as well as the spacing between adjacent phase screens, $\kappa$ is spatial frequency, $\kappa_0 = 1/L_0$, $\kappa_m = 5.92/l_0$, $L_0$ and $l_0$ is the outer and inner scales of turbulence. $R(\kappa_x,\kappa_x)$ is the Gaussian random function with the average value of 0 and variance of 1.

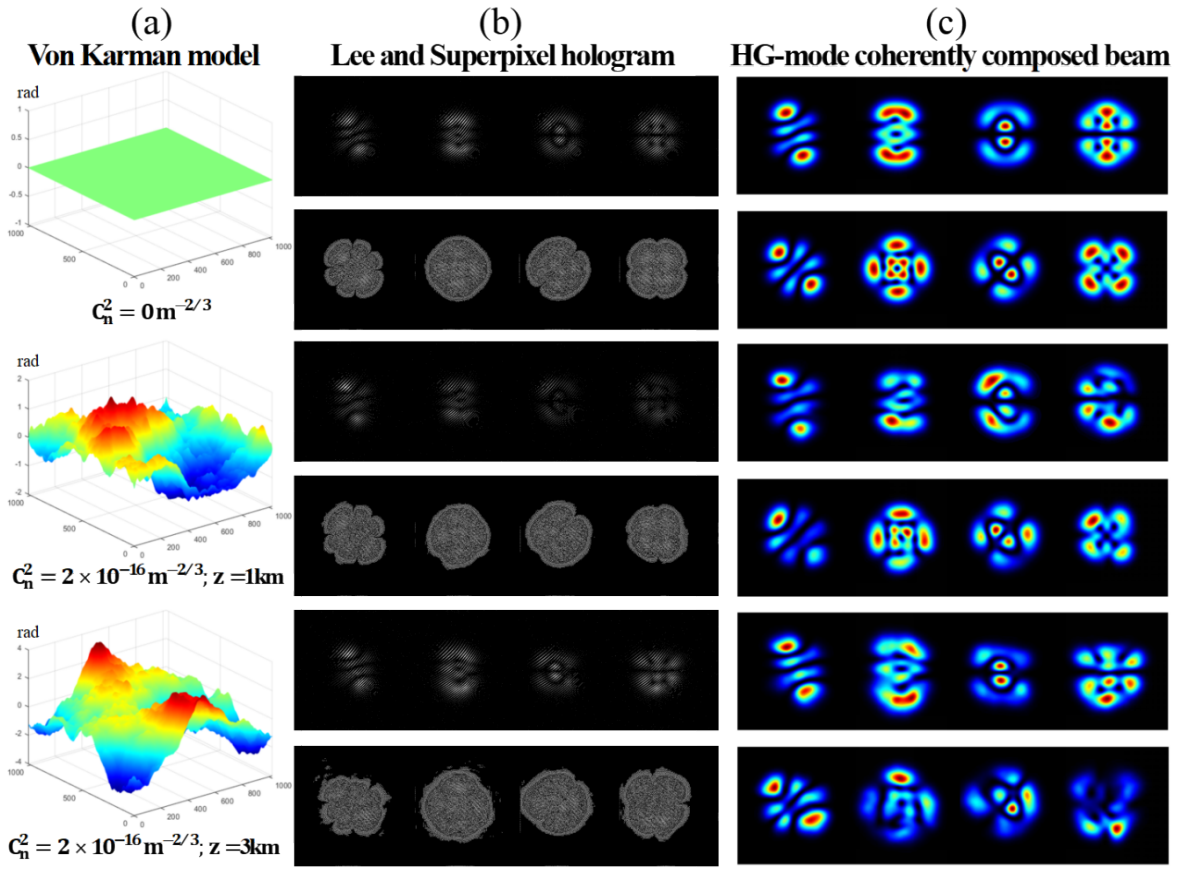

Fig. 2. HG-MCCS with additional atmospheric turbulence disturbance generated by the same DMD. (a) Random phase fluctuations are caused by atmospheric turbulence with different refractive index structure parameters and simulated transmission distance according to the Von Karman model. (b) The Lee and Superpixel holograms of single HG-MCCS and beams with different degrees of disturbance phase term. (c) The intensity distributions of the generated HG-MCCS.

The mathematical expressions of HG-MCCS with distortions from turbulence $S(x,y)$ are calculated by Eq. (10).

$$S(x,y) = E(x,y) \cdot \exp[i\varphi(x,y)] \quad (10)$$

The light field after transmission is obtained by piecewise integral simulation of the total transmission distance z according to the transmission model. Then the holographic mask on DMD can be calculated according to Eq. (6)-(8), and both the Lee and Superpixel method can be applied as shown in Fig. 2(b). After the collimated

Gaussian beam is modulated by these DMD masks, the diffracted beam is spatially filtered by a pinhole aperture to select the +1st order. The targeted HG-MCCS with corresponding distortions can be obtained in the far field as shown in Fig. 2(c). The Strehl Ratio (SR) calculated by measuring the ratio of on-axis light intensity with and without turbulence, matches well with the theoretical curve. Therefore, DMD modulation by this method can accurately simulate different turbulence levels and add different degrees of distortions to the HG-MCCS.

*2.3 Recognition of HG-MCCS based on CNN*

The recognition of the HG-MCCS patterns with distortions relays on the convolutional neural network (CNN) based pattern recognition method. As there are already lots of works focusing on beam pattern recognition with the deep learning algorithm including self-organizing map (SOM), CNN, and other network architectures, we use a similar algorithm to recognize the HG-MCCS patterns with turbulence disturbance. The AlexNet network structure adopted mainly includes five convolution layers and three full connection layers, which achieves a better recognition effect with fewer network layers compared with the subsequent proposed Network such as the GoogleNet [45,46]. Thanks to the obvious pattern feature of HG-MCCS, we can use a relatively simple network to achieve recognition, which can reduce the time consumption of the programs. The ReLU is used as the activation function to remove the negative value and keep the positive value unchanged in the convolution result, which successfully solves the gradient dispersion problem in the relatively deep network. The Dropout layer used in the fully connected layer randomly ignores a part of neurons to avoid overfitting the model. The Stochastic gradient descent with momentum (SGDM) algorithm was selected as the optimization function, and the initial learning rate and verification frequency were 0.001 and 30 respectively. Using images of 227×227 pixels as input, each mode under each turbulence transmission condition has 600 experimental samples to train the network and 100 validation samples to check the effect. With the increase in the number of iterations, the loss value becomes progressively smaller. When the iterations increase to 2500, the loss value is already close to 0, which indicates that the predicted mode gets close to the actual mode.

## 3. Experimental setup and results

*3.1 Experimental setup of the link*

The schematic diagram of the experimental setup to demonstrate the HG-MCCS-based FSO communication link is shown in Fig. 3. The fundamental mode Gaussian beam is collimated by the beam expander to cover the hologram completely. Adjusted by high reflective mirrors, the extended He-Ne laser beam is incident onto DMD at an angle of 24° relative to the normal direction. The image pixel information to be transmitted is encoded to the corresponding HG-MCCS mode. The loading of encoded information is realized by spatially modulating the beam by a binary amplitude hologram, which is calculated by the Lee method and Super-pixel method mentioned above separately. The resulting beam enters the 4f imaging system in time sequence and is spatially filtered by the aperture diaphragm. The turbulence disturbance of different intensities in the propagation process of the filtered +1st diffraction beam is simulated according to Eq.(9) and Eq.(10). After converging by the focusing lens, the image of beam intensity is collected by the CCD, whose pixels are 1280×960. In the experimental operation, the phase distortion caused by the turbulence is directly encoded on the hologram, and the HG-MCCS transmitted under different turbulence conditions are directly generated by DMD without additional SLM, so as to simplify the experimental steps. Finally, the CNN network system classifies the collected beam intensity image to identify the beam mode and converts it to the corresponding data to reconstruct the transmitted information.

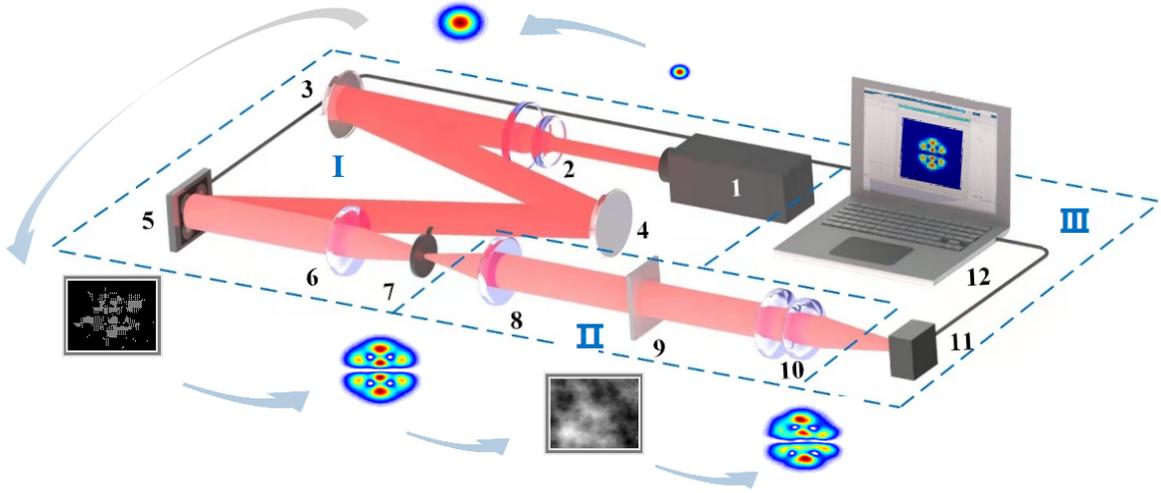

Fig. 3. Schematic of the experimental setup for information encoding/transmission/decoding using HG-MCCS. Regions I, II, and III correspond to the generation, transmission, and recognition modules of HG-MCCS respectively. 1: He-Ne laser; 2: the beam expander; 3 and 4: high reflective mirrors; 5: digital micro-mirror device (DMD); 6,8 and 10: lens with focal lengths of f6 = 100 mm, f8 = 100 mm, and f10 = 50 mm respectively; 7: aperture diaphragm; 9: atmospheric turbulence; 11: charge coupled device (CCD); 12: computer.

*3.2 Experimental results of the HG-MCCS based FSO link*

In the experiment, we used the simple two eigenmodes composed HG-MCCS to code, transmit and decode the information of colored images. With HG eigenmode order N≤6, and initial phase difference Δφ= 0 and 1/2π, and scale coefficients $\alpha_1=\alpha_2$, a total of 139 kinds of HG-MCCS are acquired. For comparison, if selecting LG-MCCS, only 40 kinds effective encoding modes at N≤6 can be acquired, and the highest mode order of 10 is needed for the 140 effective coding modes. The first 125 ones of the 139 HG-MCCS are selected for information encoding in our experiment. Fig. 4(a) shows partial examples of the first 125 HG-MCCS with different orders. The beam patterns, coding serial numbers and theory formulas of the HG-MCCS are noted. The corresponding experimentally measured images of AT distorted beam patterns are shown in Fig. 4(b), with the AT of $C_n^2 = 2 \times 10^{-16}$ m$^{-2/3}$ and z=1 km.

The beam patterns of the coding modes are clear under weak turbulence, which are easy for CNN to extract features for classification and recognition. Meanwhile, there is little difference in beam radius of all these used HG-MCCS, which is easy to be collimated and received by the optical system. The recognition accuracy of each HG-MCCS is tested for 100 times to achieve the averaged value. The average accuracy line chart and corresponding crosstalk matrix under different turbulence conditions are shown in Fig. 4(c). In the case of no turbulence disturbance, and simulated turbulence conditions of $C_n^2 = 2 \times 10^{-16}$ m$^{-2/3}$ with z=1 km, and $C_n^2 = 2 \times 10^{-16}$ m$^{-2/3}$ with z=3 km, the total average recognition accuracy of 125 modes are 99.47%, 97.52%, and 92.80% respectively. It can be seen that the overall accuracy rate of 125 modes is close to 100%, especially under the condition of no turbulence. Even under turbulence conditions, the FSO link has a good recognition accuracy. Only very individual modes under turbulence have significantly higher error rates than average, which is prone to crosstalk. A certain proportion of training samples were added to the database to enhance the recognition ability of CNN for these modes. It also should be noticed that, here, we only use a very simple CNN program architecture to achieve these results. Further improvement of the program should lead to better recognition accuracy, especially for HG-MCCS beams with heavy AT conditions.

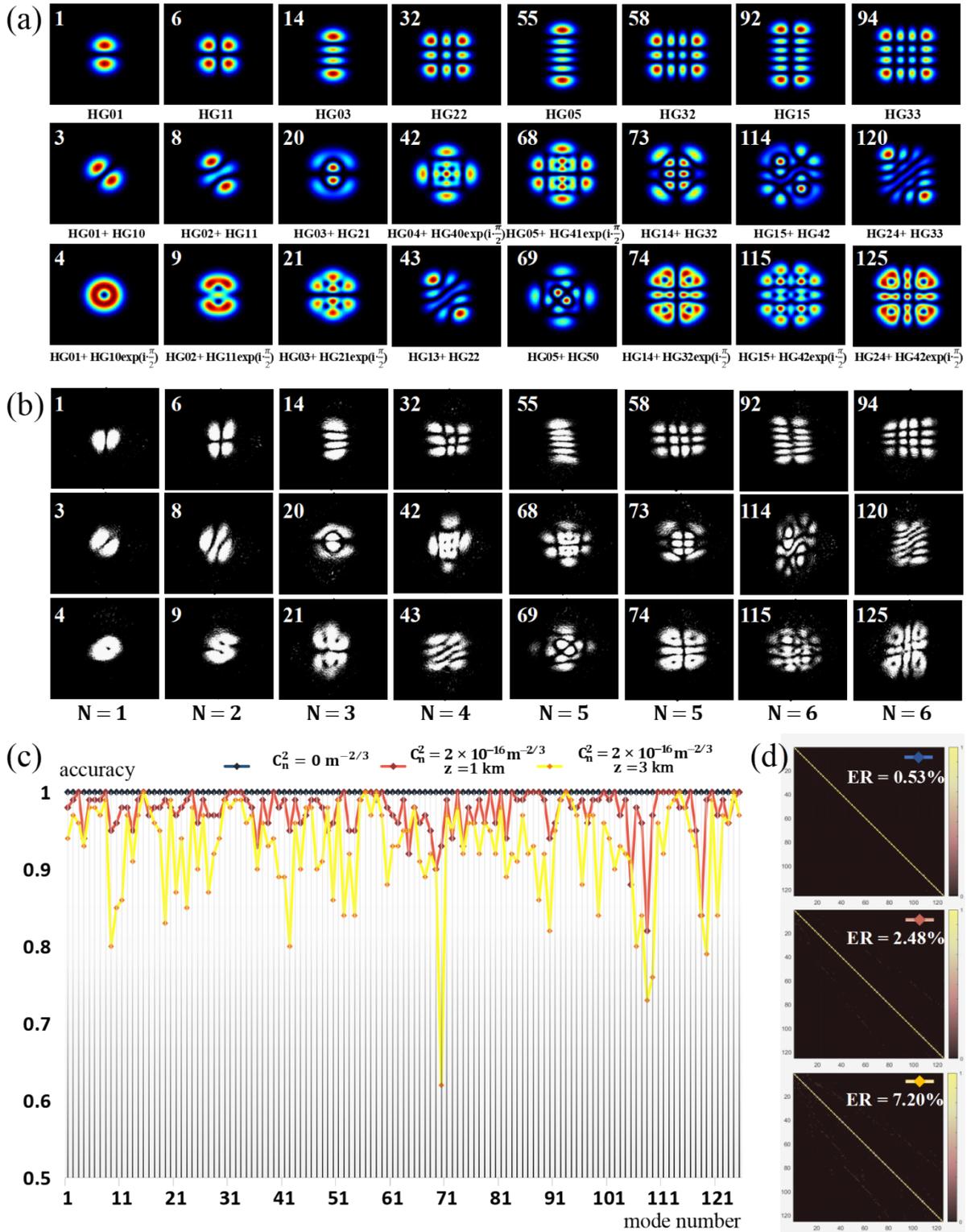

Fig. 4. The mode coding alphabet and error rate evaluation of HG-MCCS. (a) Partial examples with different orders in the encoding beams with corresponding mode serial numbers and theory formulas noted. (b) Experimental beam profiles corresponding to coding modes at the AT of $C_n^2 = 2\times 10^{-16}$ m$^{-2/3}$ with z=1 km. (c) Average accuracy line chart for 100 times recognition of each mode. (d) Crosstalk matrices of 125 kinds of HG-MCCS under different turbulence conditions.

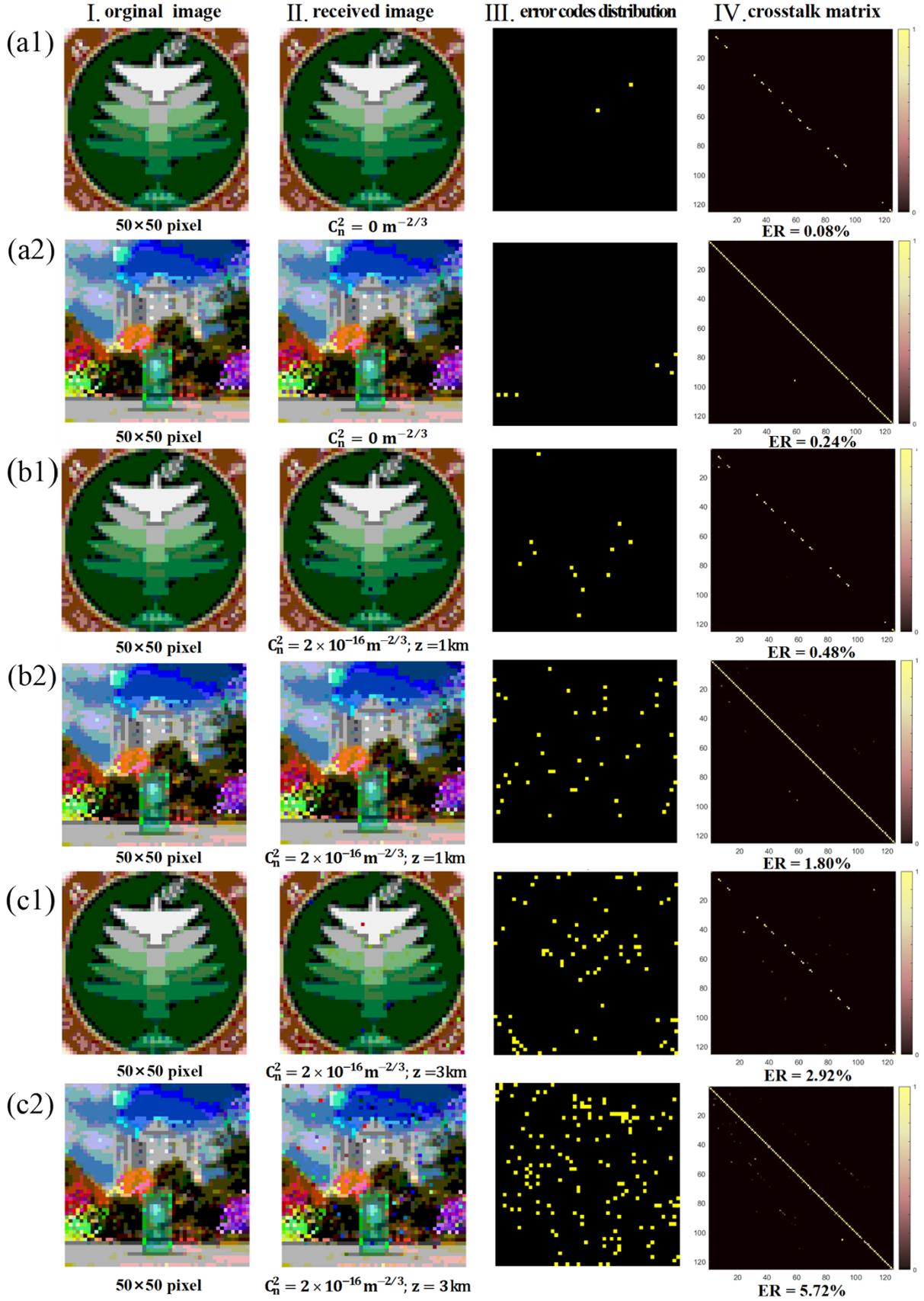

Fig. 5. Experimental performance of the HG-MCCS and CNN based FSO link. The I to IV columns are original images, received images, distribution of error codes (marked by yellow squares), and corresponding crosstalk matrices respectively. (a) Experimental results of 125-level color images transmitting without turbulence. (b) Experimental results of 125-level color images transmitting at the AT of $C_n^2 = 2 \times 10^{-16}$

$m^{-2/3}$ with z=1 km. (c) Experimental results of 125-level color images transmitting at the AT of $C_n^2 =2\times 10^{-16}$ $m^{-2/3}$ with z=3 km.

In order to further verify the performance of the link, color images with the size of 50×50 pixels are transmitted with or without simulated turbulence disturbance. The three-color dimensions RGB of the image each are divided into 5 levels, constituting a total of 125 kinds of chroma information, which are encoded by 125 kinds of HG-MCCS respectively. Even holograms corresponding to the same code are slightly different due to the random additional phases caused by simulated turbulence. The received image results under different transmission conditions including the distribution of error codes and the corresponding crosstalk matrices are depicted in Fig. 5(a)-(c). Due to the color characteristics of the transmitted school badge image, the encoding modes corresponding to some RGB values are not used, resulting in the missing of some diagonal positions of the crosstalk matrix. The bit error rate of the received school badge image without turbulence is 0.08%. In the case of simulated turbulence conditions of $C_n^2 =2\times 10^{-16}$ $m^{-2/3}$ with z=1 km and $C_n^2 =2\times 10^{-16}$ $m^{-2/3}$ with z=3 km, the bit error rates of the received school badge image are 0.48%, and 2.92%, as shown in Fig. 5(a1)-(c1). With the increase of intensity and transmission distance in turbulence, the distortion of beam intensity structures become more serious, leading to the increasing error rate of recognition. However, due to a mass of training samples under turbulence conditions added into the database, CNN has a strong ability to recognize modes with such distortions and significantly improves the accuracy rate of information transmission.

At the same time, we also transmit another building image with complete 125 colors to verify the universality of the system. And the crosstalk matrix of the building image encoded by the full alphabet is continuous on the diagonal. The bit error rates of the received building image are 0.24%, 1.80%, and 5.72% respectively under the simulated transmission conditions of non-turbulence, $C_n^2=2\times10^{-16}$ $m^{-2/3}$ with z=1 km and $C_n^2=2\times10^{-16}$ $m^{-2/3}$ with z=3 km, as shown in Fig. 5(a2)-(c2). Due to the stronger similarity between adjacent HG-MCCS coding modes, the bit error rates of images encoded by all the 125 coding modes is slightly higher than that of the images encoded by discrete coding modes. On the whole, the crosstalk occurs in few modes, and the distribution of error codes is scattered, which hardly affects the overall impression of the image. The recognition accuracy of the system is quite high, and achieves the communication function including encoding, transmission, and decoding of color images both for the case of no and weak turbulence.

## 4. Conclusion and discussions

With the extensive investigations on structured beams, their complex spatial and phase characteristics make them suitable for expanding communication capacity as high-dimensional information carriers in FSO. In this paper, we introduce HG-MCCS determined by three independent parameters including indexes of HG modes, relative initial phases, and scale coefficients to form the coding mode set. The HG-MCCS has a large number of effective divergence-degenerate modes at lower mode order, thus significantly increasing the communication capacity in a limited aperture. Meanwhile, its uniquely corresponding beam intensity distribution with evident characteristics is very suitable for the CNN decoding method based on image recognition, and CNN's generalization ability further reduces the bit error rate under weak turbulence conditions. We experimentally demonstrate an encoding and decoding FSO link based on HG-MCCS, which can realize the transmission of 125-level color images. Benefiting from CNN's fast and high-precision recognition ability and more consistent propagation characteristics between coded beams, the image transmission effect is still ideal even under the weak AT condition.

Not only does the coding number increase exponentially with the mode orders but also the three key parameters can be further utilized. In the follow-up work, the relative initial phases $\Delta\varphi_i$ and scale coefficients between modes $\alpha_i$ can be fully exploited by setting a smaller phase difference interval or adjusting the proportion of HG modes. More available coding HG-MCCS without increasing the upper limit of the mode order can further expand the communication capacity without increasing the requirements of the transmit-receive system. At the same time, information decoding with higher accuracy under different transmission conditions can be achieved by optimizing the CNN network structure or enriching training sets under different turbulence intensities. With the development of CNN, the more advanced network structure will further reduce the training time and training data volume and improve the recognition speed. Under stronger turbulence conditions, the system can be easily combined with the GS algorithm and adaptive technology for phase compensation to improve the recognition accuracy. Therefore, the communication system has a good application prospect.

**Disclosures.** The authors declare no conflict of interest.

**Data availability.** The data that support the findings of this study are available from the corresponding author upon reasonable request.